# A Physically-Based Particle Model of Emergent Crowd Behaviors

Laure Heïgeas♦, Annie Luciani♠, Joelle Thollot♦, Nicolas Castagné♠

♠ ICA Laboratory, INPG, Ministère de la Culture.
♦ Artis project-GRAVIR, CNRS-INPG-INRIA-UJF, Grenoble, France.

## Abstract

This paper presents a modeling process in order to produce a realistic simulation of crowds in the ancient Greek agora of Argos. This place was a social theater in which two kinds of collective phenomena took place: interpersonal interactions (small group discussion and negotiation, etc.) and global collective phenomena, such as flowing and jamming.

In this paper, we focus on the second type of collective human phenomena, called *non-deliberative emergent crowd phenomena*. This is a typical case of collective emergent self-organization. When a great number of individuals move within a confined environment and under a common fate, collective structures appear spontaneously: jamming with inner collapses, organized flowing with queues, curls, and vortices, propagation effects, etc. These are particularly relevant features to enhance the realism - more precisely the "truthfulness" - of models of this kind of collective phenomena.

We assume that this truthfulness is strongly associated with the concept of emergence: evolutions are not predetermined by the individual characters, but emerge from the interaction of numerous characters. The evolutions are not repetitive, and evolve on the basis of small changes.

This paper demonstrates that the physically-based interacting particles system is an adequate candidate to model emergent crowd effects: it associates a large number of elementary dynamic actors via elementary non-linear dynamic interactions. Our model of the scene is regulated as a large, dynamically coupled network of second order differential automata.

We take advantage of symbolic non-photorealistic and efficient visualization to render the style of the person, rather than the person itself. As an artistic representation, NPR reinforces the symbolic acceptance of the scene by the observer, triggering an immediate and intuitive recognition of the scene as a plausible scene from ancient Greece.

*Keywords: Crowd simulation and rendering, emergent behavior, physically-based particle model, image-based rendering.*

## 1. INTRODUCTION

Modeling crowds is currently an important field of study in Computer Graphics. Crowd simulation and rendering make virtual environments more life-like and believable. Solutions are being developed in order to address various industrial applications in entertainment, marketing, ergonomics, safety, virtual prototyping, cultural heritage, etc.

This paper presents a modeling process we developed in order to recreate a very convincing simulation of crowd movement with the simplest possible model. We focused on a type of collective human behavior called *emergent crowd phenomena*. Crowd phenomena are typical cases of collective self-organization, that is to say they are an emergent behavior. If a large number of individuals are placed inside a partially confined environment, collective structures, such as groups or flows, appear spontaneously. Our aim was to reproduce these essential structures with a straightforward approach.

The chosen case concerns an archeological site, the agora of Argos. This place was a social arena in which people came to wander, discuss, negotiate, and participate in collective events such as sporting competitions or theatrical performances. Among all the possible dynamic structures of crowd evolutions, we did not focus on those that are symbolic or cooperative, such as negotiation, discussion, cultural or social interaction, etc. but rather on those that are non-deliberative: flowing, avoiding, jamming, collapsing, etc.

The methodology we used was *first* to specify precisely the main relevant features for these types of crowd evolutions, and *second* to define the simplest generic model able to meet all these specifications. The results prove that a physically-based particle approach is particularly appropriate in this case.

To evaluate perceptively the obtained simulations with the minimal bias in interpretation, we created multiple visualizations: evolutions of parts of trajectories for flowing and whirling effects, points for jamming effects, etc.

Finally a 3D visualization was built, coating each person of the crowd by means of a non-photorealistic representation in order to obtain an interactive and symbolic rendering of the crowd.

## 2. RELATED WORKS

Numerous works have been published in the area of crowd simulation. They refer to different modeling processes, i.e. different ways for analyzing and understanding the relevant features of collective phenomena.

At the present time, crowd animations can be modeled using three main approaches:

- A kinematic approach, in which key frames and interpolations preset the animation.

- An approach based on agent systems in which agents are managed in real time by rules of behavior defined with automata.

- An approach with a particle-based system where the particles are animated in real time by the application of different forces.

For the kinematic approach, the evolutions of the displacements and of the trajectories are explicitly defined by temporal evolution functions. It attempts to produce the effects without considering their causes: it is *a phenomenological approach*. It takes advantage of the fact that every movement can be rendered, whatever it is. This method becomes time consuming when there are many characters that must avoid each other and bypass obstacles. Raupp Musse and Thalmann [20] automates the determination of trajectories for a group of characters by providing a set of Bezier curves that do not collide. The kinematic approach is pre-computed and thus totally controlled: this method is not suitable when the goal is to simulate unpredictable collective behaviors.

This type of kinematic model is called by Lantin [11] "*one-shot model*". The two other methods are *generative approaches*, describing possible causes that may produce the desired effects. Generative approaches take advantage of the fact that several complex behaviors can be synthesized with a single model; however it is often difficult to find the optimal generic model, if it exists.

Indeed, since crowd behaviors are essentially emergent, generative approaches such as agent systems or physical models are most appropriate to reproduce these kinds of phenomena.

Agent models are best adapted to model behaviors with strong individual differentiation, such as cooperative behaviors in which the actors' intentions play a significant role (collective sports, joint action, etc.).

The approaches of Thalmann et al ([20], [29]), Deviller et al [4], Donikian [5], or Tecchia, Loscos et al [26], use complex finite automata to determine actors' behaviors. These automata represent intelligent autonomous behaviors defined by sets of clever rules. Interactions between persons are modeled by symbolic rules and constraints. When there are many agents, the complexity of the modeling increases considerably. Thus, the number of agents remains usually small, typically insufficient to model large crowds. Agent-models with rules and constraints cannot be adapted to model chaotic phenomena. Dynamic and complex effects are not easily obtained by describing a set of symbolic rules and constraints. Nevertheless, Lantin [11], Fleisher [6], Logan [13] have modeled self-organizing structures, i.e. complex emergent structural effects, using similar approaches for the simulation of the growth of living organisms.

Reynolds [21] has addressed the modeling of emergent collective phenomena by agent-based systems. He extended this work by adding a steering motor force to the agent-particles [22]. The autonomous particles are classified either as engines or as autonomous characters with their own internal behavior. Goldenstein et al. [7] use a similar agent-particle system with different collision detection and path finding techniques.

In the case of collective emergent behaviors such as in crowds, the basic phenomenon is mutual, implicit, non-conscious and non-deliberative adjustment, in which global and external dynamic effects are more prevalent than steering or deliberative cooperation. In this mechanism of adjustment, collisions and avoidance are included implicitly. As in the works of Luciani [14] modeling sands, fluids and pastes, this mechanism may be simulated with physically-based particle models incorporating two elementary repulsive forces, or more generally as dynamic interactions governed by the Newtonian principle of action-reaction. This type of physical model is largely used today to simulate the phenomena of traffic jams [18] or sand heap dynamics [15].

Particle modeling has already been used by Bouvier et al [2] to simulate crowds, even though it was not used here to regulate the dynamics of avoidance and anticipation. In this work, a geometrical estimation for collision detection was used instead of a physically-based computation of forces.

In our model, the crowd is represented by a large set of simple dynamic interactions at a physical level, rather than at a level of rule-based decision-making, and this approach enables all the interactions between characters, obstacles and targets to be taken into account without increasing the design difficulty.

The contribution of the work presented here is to demonstrate that the physically-based particle system is the best candidate to design a generic model which will simultaneously (1) be the simplest model and (2) be able to produce the main emergent features of collective phenomena such as those appearing in non-deliberative crowd evolutions.

## 3. ANALYSIS OF THE DYNAMIC EFFECTS

In our approach, we do not attempt to model the phenomenon's real causes, but rather a generic process, playing the role of a *possible* cause, and capable of recreating the expected characteristic dynamic effects. Thus we need two steps of equal importance: (1) analysis; (2) modeling and synthesis. The analysis is done by observing the natural phenomenon, as "a physical phenomenon", without any symbolic interpretation. Its aim is to list the specifications describing the relevant qualitative and quantitative dynamic properties of crowd behaviors. The modeling then consists of finding the simplest physically-based particle model able to satisfy all the specifications identified in the analysis stage.

### 3.1 Case study

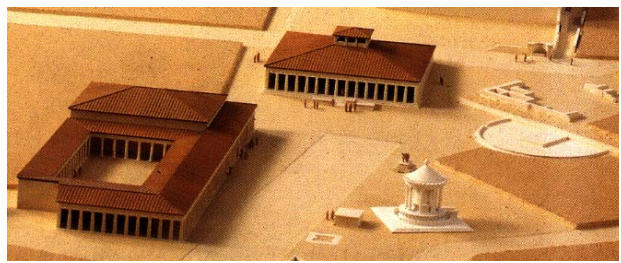

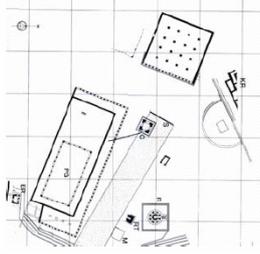

**Figure 1:** *Agora of Argos during the reign of Emperor Hadrian (117-138 A.D.): [1] [19]*

Our analysis mainly refers to our case study, which is the agora of Argos (fig. 1). This agora is a locus in which two kinds of collective phenomena took place: interpersonal interactions (such as group discussion and negotiation), and global collective phenomena (such as large group displacements and collective events). As explained in section 2, the first type of behaviors, cooperative behaviors, is outside the scope of this work.

Furthermore, this choice of the agora of Argos is also motivated by the need to propose to archaeologists a way to validate assumptions about life in this kind of ancient place, such as assumptions regarding evolution of density, plausible repartitions, dynamic arrangements, etc.

### 3.2 Crowd definition

In this work, we adopt the point of view of sociologists ([12], [23], [16]) who show that a crowd is not, or is not only a great number of agents in a common space. P. Livet [12] defined an alternate point of view to the well-known Durkheim's approach. He introduced the idea that a crowd is not an object (for example a large number of differentiated agents) that presents different behaviors, but a more phenomenological entity defined as a class of behaviors. It is in that sense that Livet talks about "virtual community": a community is virtual because it is more an effect than a tangible thing. Following this point of view, we will define a crowd as "a set of quasi-similar objects or persons, inside a common environment, following a common fate (for example, flowing), and *exhibiting emergent collective common features*".

That is an alternate point of view to those underlying computer graphics models, mainly based on the modeling of the crowd as a complex object, and exhibiting complex behaviors explicitly programmed.

### 3.3 Specification of crowd dynamics

The analysis of the dynamic structures that appear in crowd movements is one of the interests of modern social science [12, 23, 29, 16]. Our analysis is based on such social observations. In the movements of crowds, two essential dynamic structures appear: jamming, i.e. the effects found in traffic jams, and flowing, i.e. processions.

Jamming is characterized by the way in which an accumulation increases, stabilizes and disappears. Regarding flow effects, emphasis is not laid on individual displacements but on the movement of the whole crowd as people go from one zone to another in a constrained manner. Among the collective phenomena, some queuing, curling or collapsing effects can be observed: some persons are temporarily trapped inside a flow, constrained to follow the same path or to form a local whorl, and others are carried away by the crowd.

### 3.4 Obstacles definition

We consider both fixed obstacles, i.e. buildings and moving obstacles, i.e. people. From a dynamic point of view, the main task is the coordination between the individual trajectories relative to the other components of the scene. The core effect to be synthesized accurately is *obstacle avoidance*. Physically speaking, the dynamics of avoidance are not limited to static non-penetration and collision detection, but also represent how an obstacle is bypassed. Thus we can define an apparent physical volume that corresponds to a zone in which something can enter, but with greater or lesser difficulty.

For physical objects, the apparent volume is different in a dynamic situation, i.e. situation of displacement, or in a static one. In the static case, it is strictly their material volumes that are impenetrable, such as supports. In the dynamic case, people walk around these objects at a reasonable distance, and thus their apparent volumes become larger. It is the same for some zones; even if their access is not strictly prevented, they are not accessible culturally (as race tracks), or physiologically, (as the zones in a hot sunlight). All these objects may be treated as obstacles that are partially or fully impenetrable [10].

We can use the same notion, often named *psychophysical volume*, for people. For example, this volume varies according to the place where the person is: this volume will be bigger on a large place than on a small alley.

Thus, each component of the scene will be considered as an obstacle to be avoided and represented by its *apparent volume*.

## 4. SYNTHESIS OF THE DYNAMIC EFFECTS

### 4.1 Principle

The aim of the following dynamic model is to simulate the assumptions made during the analysis of crowd dynamics.

The basic idea consists in choosing the minimal physical model for individuals and their interactions with the environment and other individuals. Morphological and anthropomorphic differentiations between individuals are not relevant for the collective movements. Thus, we can assume, as a starting point, that individuals can be modeled as particles on a two-dimensional ground and that non-mobile objects can be modeled as sets of fixed particles. The interactions between individuals are represented by physically-based forces exerted between each pair of particles.

### 4.2 Interaction

The basic interaction among the crowd has to represent the dynamic way in which characters avoid obstacles. As seen in section 3.4, any obstacle can be considered as a psychophysical volume. The interaction law we depict in this section was modeled according to this notion.

In order to obtain a general and extensible formalism, we choose the most generic pair-wise interaction function able to recreate all the dynamic behavior of pairs of particles. As discussed in [14, 17, 27, 28], it is composed of two terms: a non-dissipative term (or potential term) and a dissipative viscous term. The most generic expression of non dissipative potential term is represented by the family of functions analytically given by

$$F(D) = -\frac{a}{D^n} + \frac{b}{D^m}$$

where $D$ is the distance between interacting particles and $a$, $b$, $m$ and $n$ some parameters (see Fig. 2). Note that the Lenard-Jones function is represented by one of these functions. However, because of the analytical formulation, the parameters are not easy to manipulate in order to control the physical type and size of behavioral zones: attraction and repulsion slopes, thresholds between attractive and repulsive parts [9].

In our model, we simplify the functions of this family with piece-wise linear functions. This allows both optimization of the computation cost and direct manipulation of relevant parameters, i.e. the distance thresholds and the slopes of each piece. Moreover we represent the dissipative term similarly to the non-dissipative potential term by a finite state automaton conditioned by the same distance thresholds (see Fig. 2). Thus, the discreet generic expression for the interacting force between two particles is defined in our model by:

$$\begin{aligned}
&\text{if} \quad D < D_1 \quad &\text{then} \quad \vec{F} = (K_1 D + Z_1 V)\vec{u} \\
&\text{if} \quad D_i < D < D_{i+1} \quad &\text{then} \quad \vec{F} = (K_{i+1} D + Z_{i+1} V)\vec{u} \quad \text{for} \quad i = 1,2 \\
&\text{if} \quad D_3 < D \quad &\text{then} \quad \vec{F} = \vec{0}
\end{aligned}$$

where $D$ is the distance between the particles and $V$ is the norm of their relative velocity, and $\vec{u}$ the unit vector between the two particles. The three thresholds $D_i$, stiffnesses $K_i$ and viscosities $Z_i$ (i=1,2,3) regulate explicitly the behavior of each spatial zone.

The figure 2 shows how we approximate the continuous non-dissipative term (left) by a piecewise linear function, and presents the final state automata we use for the computations (right).

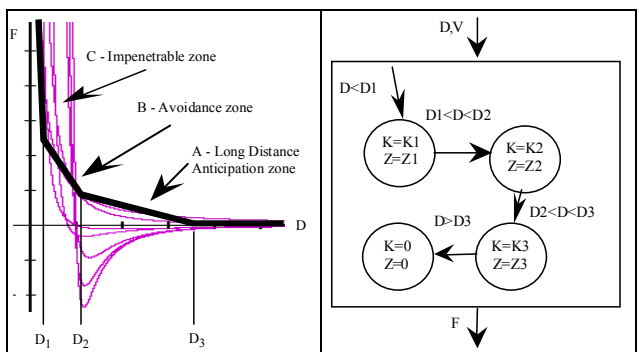

**Figure 2:** *Interaction for obstacle avoidance*
*Left: Family of elastic interaction functions
and piecewise linear approximation
Right: final state automata used for computation*

The first zone A is the anticipation long-distance zone with a low stiffness. The middle zone B with a medium stiffness regulates the beginning of the repulsive force. The last zone C regulates the effective repulsive force: the slope, i.e. stiffness, is sufficiently high compared to the velocity of a free particle (around 1.5 meters per second in our case) to make the zone impenetrable. By adapting the value of the viscosity for each zone, it is possible to regulate the velocity behavior. For example, a high viscosity in zone A slows the person down significantly.

These functions model the interaction between individuals and obstacles according to three psychophysical spheres. The parameters can vary according to each pair of particles in interaction. Thus the system exhibits a large variety of avoidance behaviors and enables us to control their style.

In the following sections, we focus on the persons' and buildings' interactions by specifying the parameters of the interaction functions.

### 4.3 Characters' interactions

The crowd's density in a given zone is determined by the interactions between the individuals and more precisely by the values of the thresholds and of the associated stiffness and viscosity, according to Figure 2. Thus, in a large place, these spheres can be very broad and rather stiff. The population will then tend to spread over the place. In a very constrained space like a narrow alley, the individuals will agree to compress themselves: the spheres radii have to be smaller with lower stiffness in order to increase the compressibility of the system and with higher viscosities for a larger damping.

In our case, the agora is a large place and according to historical data, this place was rarely overcrowded (see Section 3.2). Based on an historical estimation [19], there was roughly an average of 15 m2 per person, which corresponds to a circle with a radius of about 2 meters.

These observations allow us to determine the various thresholds. Thus, the impenetrable threshold is at 1 meter, the psychophysical avoidance threshold can begin at around 3 meters and the anticipation threshold at around 5 meters.

In fact, for more realism, a crowd needs little variations among the individual behaviors. All the individuals do not approach the obstacles in the same way and these distinctive signs are very significant in the crowd modeling. Thus the previous thresholds, their corresponding stiffness and viscosity are uniformly distributed between minimal and maximum values.

### 4.4 Buildings as obstacles

We will not give a geometrical representation of the buildings walls, but rather a dynamic representation of the interaction between buildings and people, i.e. the psychophysical representation of the buildings: the particles positions representing a building do not depend on its geometrical border but on the mental representation of a building as an obstacle to avoid.

Buildings are modeled as sets of fixed particles interacting with people. These particles are assembled without any automatic computation and we have to choose individually the positions and thresholds associated to each particle. The method used to arrange this set of particles is based on the way people approach obstacles. If a wall blocks the advance of a person, the effect of anticipation increases with the size of the wall: the larger the wall, the more significant the anticipation and the earlier the person acts to avoid the wall.

Moreover, a person circumvents a building, not by following the walls, but by going directly toward the corner. It appears as if the person is more strongly repulsed by the walls than by the corners. Based on these observations, square obstacles are modeled in our model by a single mass with a circular zone of interaction.

A single particle is not sufficient to model a rectangular building, because its circular interaction would be too large. So, we use several particles in order to create a kind of envelope around the building. In Fig. 3, the circles represent the spatial singularity of the interaction between buildings as obstacles and particles representing people.

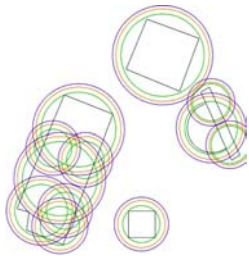

**Figure 3:** *Buildings as obstacles.*

### 4.5 Crowds injectors and targets

The characters composing the crowd enter the place from various injector sources.

An injector is defined by the number of persons who will come from it and its intensity of flow, that is the number of people generated per second. These injectors allow simulation of various crowd behaviors, such as expansion effects when people come from a small street into a larger space (regular and continuous effects), or procession effects when all the people are injected nearly at the same time. In the model, each person has a target location, set for the entire simulation. The character's target is modeled with an attraction force, i.e. a visco-elastic interaction between the character and its target.

## 5. VISUALISATION

The visualization plays the role of the tool of validation. The validation could not be done by direct comparison, since our case study is an antique site. The principal criterion of validation is plausibility, which is also a valid criterion in the general case. We carried out three types of visualizations. Each one underlines different behaviors; thus their combination permits a greater acuity in observing the phenomena generated by our model. We validated the trajectories obtained through 2D visualizations, and we built a 3D rendering in order to present it to a larger audience.

### 5.1 2D visualization of trajectories

The phenomena relating to the movements are better observed in 2D because they are related only to the characters trajectories on the ground.

Although the modeling of the characters does not refer to intelligent rules, either cultural or learned, we observe much more than simple obstacle avoidance. Indeed, advanced behaviors appear although not explicitly computed. These include approaches with hesitation, accelerations after a strong constraint, or dynamic figures similar to jamming, linear or whirling flows.

#### 5.1.1 Disc visualization

In a first representation, people are represented by simple discs (see Fig. 5). This enables us to see efficiently how people are spread over the scene and which spatial arrangements and dynamics of avoidance are obtained.

In the first part of the simulation, we can see people bypassing the buildings and jamming near them, while less constrained people in the center of the place walk easily. We can also follow persons carried away by a flow of people. This behavior corresponds well to what we can observe in a crowd: inside the crowd, people temporarily lose their individuality. Once free of these constraints, the characters escape and accelerate. In the second part of the simulation, there are more people. They are thus more constrained and an important structure of flows occurs: constrained persons form a queue. We can also observe that some persons jam, stop for a while and then spread out.

#### 5.1.2 Trajectory visualization

It is significant to visualize the trajectories of the persons in order to better perceive spatial and temporal phenomena that emerge within the crowd. We represent a trajectory with a line joining the previous person's positions (for a given time interval). We observe the marks of two persons avoiding each other, face to face, by walking round softly (see Fig. 4).

Other macroscopic phenomena can appear, such as the formation of a volute when several flows of the crowd meet (Fig. 6). This type of phenomenon is due to viscosity effects, part of the modeled interaction between elementary components [14].

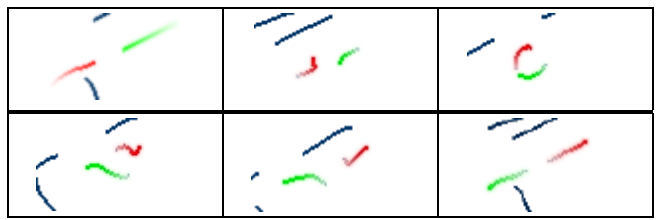

**Figure 4:** *Avoidance of two persons passing face to face.*
*Upper: they bypass each other decelerating.*
*Lower : they hesitate with small local oscillations of trajectory and then accelerate.*

### 5.2 3D visualization

Based on the trajectories obtained with our model, we built a 3D real-time visualization (Fig. 7), allowing a user to see the scene as if he were walking in the agora. This visualization gave us a powerful tool for evaluating the crowd sensation, and also gave us the possibility to present our work to a large audience such as in a museum.

Two kinds of visualizations are possible in order to render the crowds realistically: (1) 3D conventional representations and (2) optimized (3D to 2D) representations, such as bill-boarding techniques widely used in computer games. Two reasons led us to prefer the second:

(1) 3D rendering of about one thousand animated persons could not be executed easily in real time with sufficient interactivity with the user.
(2) The chosen method allows introduction of an artistic stylization, which can deeply reinforce the acceptance of the scene by making it immediately recognizable, in our case, as an antique Greek scene.

### 5.2.1 Rendering

In order to render walking people by coating the 2D trajectories obtained with the physical model, we used the image-based rendering technique of Tecchia et al. [25]. This method is based on a positional and directional discretization of avatars and on the mapping of the obtained 2D representation at the adequate step of the animation.

In order to minimize geometrical complexity, each particle, that is, each human, is represented with a single adaptive impostor. The smooth rendering of a walking human is pre-computed in the chosen rendering style and decomposed in a set of textures viewed from different viewpoints and at different times. Appropriate images for the impostors are then selected for each frame depending on the viewpoint position.

In their work, Tecchia et al. had produced as a final result very realistic images, by combining texture compression and multi-pass rendering. In our case, however, realism was not desirable. On the one hand, archeologists do not have sufficient details to reconstruct exactly the agora of Argos; and on the other hand, antique Greek clothing is easily recognizable.

Thus, an artistic or stylized rendering was more appropriate for a visualization of Argos: it brings sense to the visualized phenomenon, gives a special atmosphere to the scene, stimulates the imagination of the viewer and invites him to project himself into the universe of the ancient city. In cooperation with archaeologists and artists, we have chosen an artistic stylization in ancient Greek style, which corresponds to the traditional media used in archaeological drawings.

The rendering of the environment (buildings, terrain) uses real-time non-photorealistic rendering techniques [8, 24]: silhouette edges and creases rendering and dynamic paper, as described in Cunzi et al. [3].

### 5.2.2 Evaluation

The perception of plausibility already observed in 2D visualization of the trajectories is strongly accentuated with 3D visualization of the scene (Fig. 7; animations, see http://www-acroe.imag.fr).

As a minor limitation, the rendering of quasi-immobile particles (particles that reached their key target or that are temporarily blocked inside the crowd) is non-optimal. In these cases, the 2D avatars tend to turn about and to wander, hesitating over the direction to follow. This is due to the simplicity of the detection of direction along trajectories, which is based on a direct measure of the velocity of the particle. Possible improvements would be to provide a better method for direction detection (use of a threshold for velocity, filtering, etc.) or, perhaps more interesting, to design a more complex physical model of the individuals, whose representation is isotropic at the present time.

The visualization process is computed in real-time. The efficiency of the rendering algorithm was evaluated on a PC (Athlon XP 1700, 1.5 GHz, 900 Mb RAM) with a NVidia GeForce 4 Ti graphic card. We obtained a frame rate of 160 Hz for a crowd with 300 persons, each figure rendered with 400 animated textures of 64 x 128 (16 points of view around each person and 25 images by step). At the present time, the number of persons is limited to 2000 in order to maintain the frame rate at 25 Hz. For this evaluation, the simulation itself was pre-computed.

## 6. CONCLUSIONS

Taking into account physically-based interaction between pairs of individuals and between individuals and obstacles allows the reproduction of complex and various dynamic behaviors, such as jamming and flowing patterns visible in real crowds. The method allows the designer to create the necessary and sufficient conditions from which auto-organized structures emerge from a set of confined and moving individuals. These dynamic structures are typical signatures of such collective phenomena.

As shown in the various visualizations (Fig. 4, 5, 6 and 7 – see http://www-acroe.imag.fr for the animations), the method plausibly renders the relevant effects of interaction between human flows: queuing, trajectory deviations, velocity variations, local curls, etc., during the displacement of the persons. Finally, Figure 7 shows 3D visualization and rendering with a Non Photorealistic Rendering method.

Two possible extensions of the model are envisaged.

First, we would study an anisotropic model of each individual, introducing front/back and right/left differentiation with a local system of particles to represent each individual. Among other interests, this would avoid the problem previously cited of instability in the orientation of quasi-immobile avatars during the 3D rendering process.

Second, we would increase the number of key-targets for each character, in order to apply physical modeling to the study of the issues of path planning and path following, such as in [10].

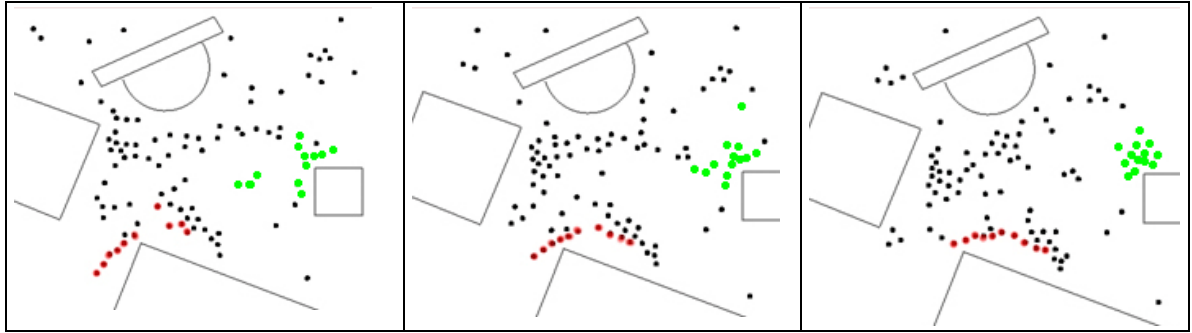

**Figure 5:** *The red discs form a queue and the green ones jam together.*

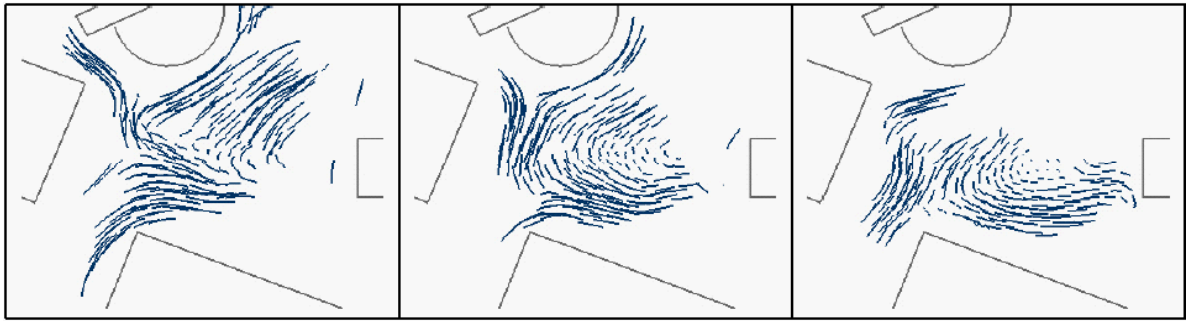

**Figure 6:** *Formation of a volute when several flows of crowd meet.*

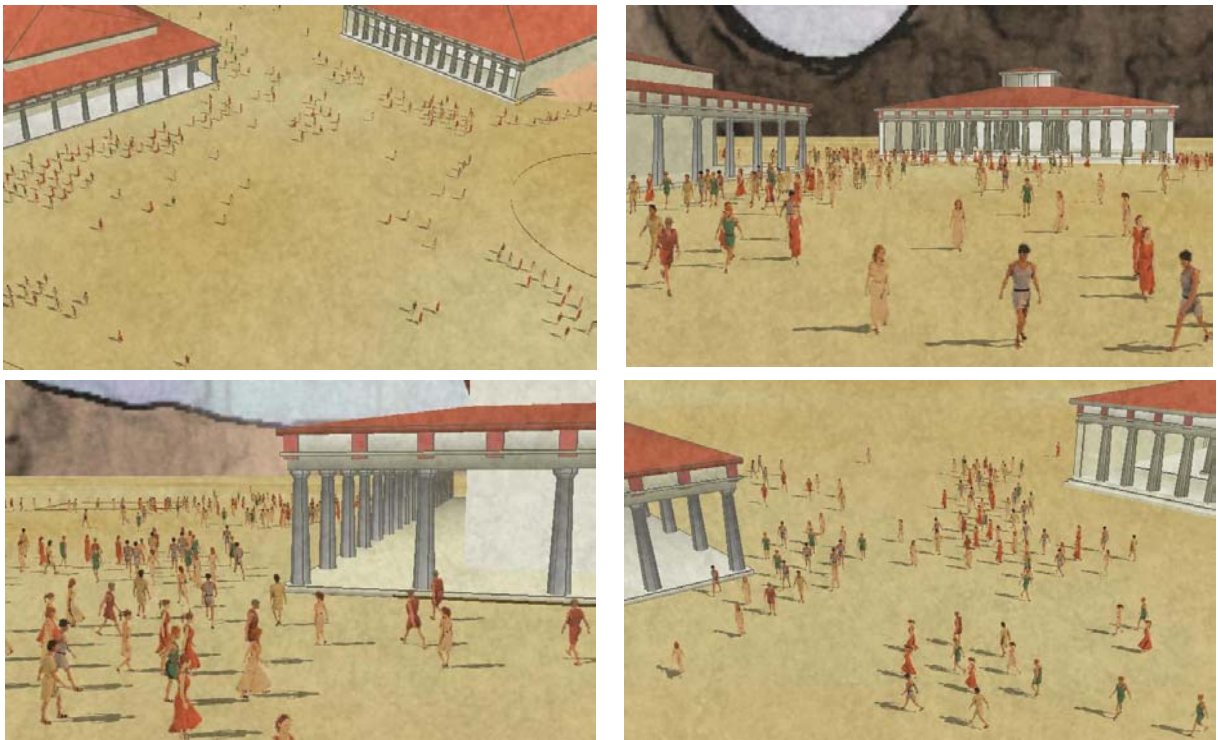

**Figure 7:** *3D real-time and stylized rendering.*

**Acknowledgements**

This research was funded in part by the INRIA Action de Recherche Cooperative ARCHEOS (http://www-sop.inria.fr/reves/Archeos) and by the French Ministry of culture, Music Department.

Some of the models shown here were created in A|W Maya by B. Ruiz and M. Negrel, students of the Ecole Superieure de Realisation Audiovisuelle in Nice.

Many thanks to the artist Isabelle Rosaz for her characters 3D models, and to Peter Torvik for his precious corrections.